\documentclass[pre,aps,twocolumn,superscriptaddress,showpacs]{revtex4-1}
\usepackage{amssymb}
\usepackage{epsfig}
\usepackage{amsmath}
\usepackage{subfigure}
\usepackage{graphicx}
\usepackage{textcomp}
\usepackage{url}
\usepackage{float}
\usepackage{color}
\usepackage{cases}
\usepackage[normalem]{ulem}

\usepackage{enumitem} 

\usepackage[colorlinks=true, urlcolor=blue, anchorcolor=blue, citecolor=blue,filecolor=blue,linkcolor=blue,menucolor=blue
]{hyperref}

\usepackage{color}
\definecolor{Blue}{rgb}{0.00, 0.00, 0.80}
\definecolor{Red}{rgb}{0.80, 0.00, 0.00}
\definecolor{Green}{rgb}{0.00, 0.50, 0.00}

\usepackage{fixmath}
\newcommand{\vect}[1]{\mathbold {#1}} 

\usepackage{epsfig}
\usepackage{textcomp}
\usepackage{epstopdf}

\newcommand{\nn}{\nonumber}
\newcommand{\be}{\begin{equation}}
\newcommand{\ee}{\end{equation}}
\newcommand{\bea}{\begin{eqnarray}}
\newcommand{\eea}{\end{eqnarray}}

\begin{document}

\title{Full distribution of the number of distinct sites visited by a random walker in dimension $d \ge 2$}

\author{Naftali R. Smith}
\email{naftalismith@gmail.com}
\affiliation{Racah Institute of Physics, Hebrew University of Jerusalem, Jerusalem 91904, Israel}


\begin{abstract}

We study the full distribution $P_M(S)$ of the number of distinct sites $S$ visited by a random walker on a $d$-dimensional lattice after $M$ steps.
We focus on the case $d \ge 2$, and we are interested in the long-time limit $M \gg 1$.
Our primary interest is the behavior of the right and left tails of $P_M(S)$, corresponding to $S$ larger and smaller than its mean value, respectively.
We present theoretical arguments that predict that in the right tail, a standard large-deviation principle (LDP)
$P_{M}\left(S\right)\sim e^{-M\Phi\left(S/M\right)}$
is satisfied (at $M \gg 1$) for $d\ge2$, while in the left tail, the scaling behavior is
$P_{M}\left(S\right)\sim e^{-M^{1-2/d}\Psi\left(S/M\right)}$,
corresponding to a LDP with anomalous scaling, for $d>2$.
We also obtain bounds for the scaling functions $\Phi(a)$ and $\Psi(a)$, and obtain analytical results for $\Phi(a)$ in the high-dimensional limit $d \gg 1$, and for $\Psi(a)$ in the limit $a \ll 1$ (describing the far left tail).
Our predictions are validated by numerical simulations using importance sampling algorithms.

\end{abstract}

\maketitle

\section{Introduction}

A standard way to quantify the territory covered by a random walker (RW) is by the number of distinct sites $S$ that it visits after $M$ steps. 
This may be used, e.g., to model the size of a habitat of an animal, or to understand the diffusion of defects in a crystal \cite{Vineyard63}.
The typical behavior of $S$ in the long-time limit $M \gg 1$ has been extensively studied. This behavior strongly depends on whether the RW is recurrent, i.e., if it is guaranteed (with probability $1$) to return eventually to its starting point. It is well known that a RW on an infinite regular lattice is recurrent in $d>2$, and non-recurrent for $d \le 2$.
The mean number of visited sites grows linearly $\left\langle S\right\rangle \sim M$ for non-recurrent RWs and sublinearly for recurrent RWs.

Considerable research has been dedicated towards understanding the behavior of typical fluctuations of $S$, which behave highly nontrivially. Indeed, although $S$ may be written as the sum of Bernoulli random variables which indicate whether each site on the lattice has been visited, these random variables are highly correlated. This makes the problem both interesting and challenging, and results, in general, in the lack of a central limit theorem for $S$, i.e., typical fluctuations of $S$ in general do not follow a Gaussian distribution \cite{JP72, WCH97, WH97, KMS13}, and the scaling behavior of the variance of $S$ with $M$ is nontrivial \cite{JP70, Torney86, MS24}.
However, the full distribution of $S$, including its large-deviation tails, has not been understood so far, except in dimension $d=1$ where the problem simplifies considerably and has been solved exactly \cite{MS24}.

In a broader context, large deviations in stochastic systems have attracted growing interest from statistical physicists over the last few decades, as their understanding often reveals interesting features related to the nonequilibrium behavior of the system \cite{DZ1998, Hollander2000, Hugo2009, Hugo2018}. The study of large deviations is motivated by many realistic applications where rare events have dramatic consequences. Examples of such events include earthquakes, stock-market crashes, pandemic outbreaks, heatwaves etc.
In particular, large deviations of sums of a large number of correlated random variables is an interesting and challenging field of study, and much research has been devoted to it in several contexts, including linear statistics in random matrices, quantum gases and Riesz gases \cite{MNSV09, SLMS20, SLMS21, VMS24, LS25}, fluctuations of long-time averages in discrete-time chaotic dynamics \cite{Smith22Chaos, GCP23, Monthus24Chaos, Lippolis24, Monthus25Pelikan, DS25} and more. 
In the context of the current work, understanding the large deviations of $S$ may be useful in applications to models that are similar to the forager model \cite{BR14}, in which the RW depletes resources that are available at each of the lattice sites. Then an unusually small value of $S$ could be linked to the death of the forager due to insufficient resource consumption.

In this work, we study the large-deviation tails of the distribution of $S$. We focus on the long-time limit $M \gg 1$ and on dimension $d>1$.
The rest of the paper is organized as follows. In Sec.~\ref{sec:Model}, we give the precise mathematical definition of the model and of the observable of interest, and present a brief summary of our main results. Our results for the right and left tails are derived in Sec.~\ref{sec:LargeDeviations}. In Sec.~\ref{sec:discussion} we discuss our main findings and outline some possible directions for future research. 
In Appendix \ref{app:Numerics} we give details of our numerical method for computing $P_M(S)$.

\section{Model and summary of main results}
\label{sec:Model}

The model we study is that of a discrete-time RW  on a $d$-dimensional hypercube lattice. We denote the position of the RW at time $t$ by
$\vect{x}_t$, where $t=0,1,\dots,M$. At each time step, the RW travels a unit distance in one of the $d$ dimensions, i.e.,
\be
\vect{x}_{t+1}=\vect{x}_{t}+\vect{\xi}_{t} \, ,
\ee
where the step $\vect{\xi}_{t}$ is chosen uniformly from one of the $2d$ vectors
$\left\{ \pm\hat{\vect{e}}_{1},\dots,\pm\hat{\vect{e}}_{d}\right\} $,
and $\left\{ \hat{\vect{e}}_{1},\dots,\hat{\vect{e}}_{d}\right\} $ is the standard (orthonormal) basis of $\mathbb{R}^d$.
We assume that the RW is initially at the origin, $\vect{x}_{0} = 0$.
The number of distinct sites visited (which is sometimes referred to as the range of the RW) is defined as the size (i.e., cardinality) of the set of points visited by the RW after $M$ steps,
\be
S=\left|\left\{ \vect{x}_{0},\dots,\vect{x}_{M}\right\} \right| \, ,
\ee
so that $1 \le S \le M+1$.

It is known that the mean number of distinct sites visited behaves as
\be
\label{meanS}
\left\langle S\right\rangle \simeq\begin{cases}
E_{d}M^{d/2}, & d<2\,,\\[2mm]
\frac{\pi M}{\ln M}\,, & d=2\,,\\[2mm]
E_{d}M\,, & d>2\,,
\end{cases}
\ee
where the numerical constants $E_d$ are known analytically \cite{Vineyard63, MS24}, and in particular,
\bea
E_1 &=& \sqrt{8/\pi} \, ,\\[1mm]
E_3 &=& \frac{1}{3\pi^{3}}\int_{0}^{\pi}\int_{0}^{\pi}\int_{0}^{\pi}\frac{dudvdw}{3-\cos u-\cos v-\cos w}\nn\\[1mm]
&=&\left(18+12\sqrt{2}-10\sqrt{3}-7\sqrt{6}\right) \nn\\[1mm]
&\times& \left[\frac{2K\left(\left(2-\sqrt{3}\right)^{2}\left(\sqrt{3}-\sqrt{2}\right)^{2}\right)}{\pi}\right]^{2} \nn\\[1mm]
&=&0.65946\dots,
\eea
where $K(m)$ is the complete elliptic integral of the first kind \cite{EllipticKWolfram} (see \cite{Watson39} for the calculation of the triple integral that gives $E_3$).
Subleading corrections to the long-time behavior \eqref{meanS} of $\left\langle S\right\rangle$ and extensions to other types of lattices have also been obtained \cite{MW65}.

Fluctuations of $S$ are more difficult to analyze since they are affected by autocorrelations of the position of the RW at different times, and these correlations decay slowly (as power laws) in time. However, the long-time behavior of the variance of $S$ is known in any dimension \cite{JP70, Torney86}
\be
\label{VarS}
\text{Var}\left(S\right)\simeq\begin{cases}
V_1 M, & d=1\,,\\[2mm]
\frac{V_{2}M^{2}}{\ln^{4}M}\,, & d=2\,,\\[2mm]
V_{3}M\ln M\,, & d=3\,,
\end{cases}
\ee
where $V_1 = 0.2261\dots$, $V_2 = 16.768\dots$ and $V_3 = 0.21514\dots$.
In fact, typical fluctuations of $S$ (and other related observables) were shown to follow universal laws in the long-time limit: At $M \gg 1$, $S$ may be written as 
\be
\label{SUniversal}
S\simeq\begin{cases}
\sqrt{M} \, \eta_1, & d=1\,,\\[2mm]
\left\langle S\right\rangle + \frac{M}{\ln ^2 T}\, \eta_{2}\,, & d=2\,,\\[2mm]
\left\langle S\right\rangle+\sqrt{M\ln M}\,\eta_{3}, & d=3\,,
\end{cases}
\ee
where the $\eta_d$ are universal distributions that do not depend on $M$ \cite{JP72, WCH97, WH97, KMS13}.

However, much less is known about large deviations of $S$, especially in dimensions $d > 1$.
The goal of this paper is to fill this gap, i.e., to understand the long-time behavior ($M \gg 1$) of the large-deviation tails of the distribution $P_M(S)$ of $S$, in dimension $d>1$. Based on upper and lower bounds that we are able to obtain analytically, we conjecture scaling behaviors for the right ($S > \left\langle S\right\rangle$) and left ($S < \left\langle S\right\rangle$) tails of $P_M(S)$. Numerical simulations that we performed using the importance sampling algorithm show good agreement with our conjecture, and enable us to numerically compute the associated scaling functions.

Let us briefly summarize our main results. We find that the right (for $d \ge 2$) and left (for $d > 2$) tails of the distribution are described by the LDP \eqref{LDPRightTail} and anomalously-scaled LDP \eqref{LDPLeftTail} respectively. This difference in the scaling reflects a strong asymmetry of the distribution $P_M(S)$, such that large deviations with $S < \left\langle S\right\rangle$ are far more likely to occur than those with $S > \left\langle S\right\rangle$.

\begin{figure*}[ht!]
\includegraphics[width=0.32\textwidth,clip=]{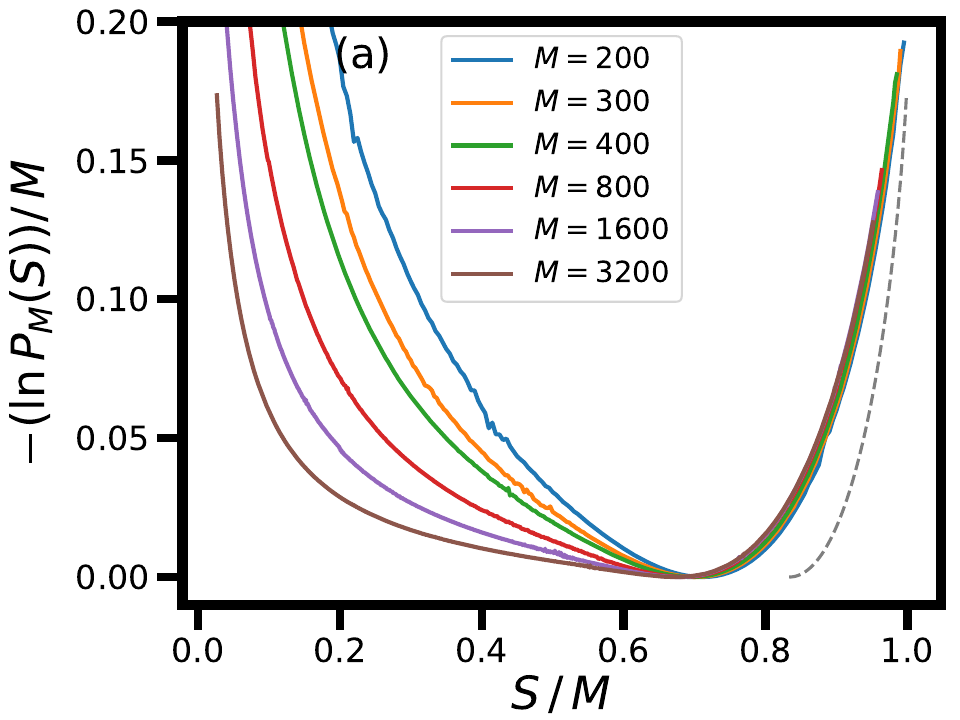}
\hspace{1mm}
\includegraphics[width=0.32\textwidth,clip=]{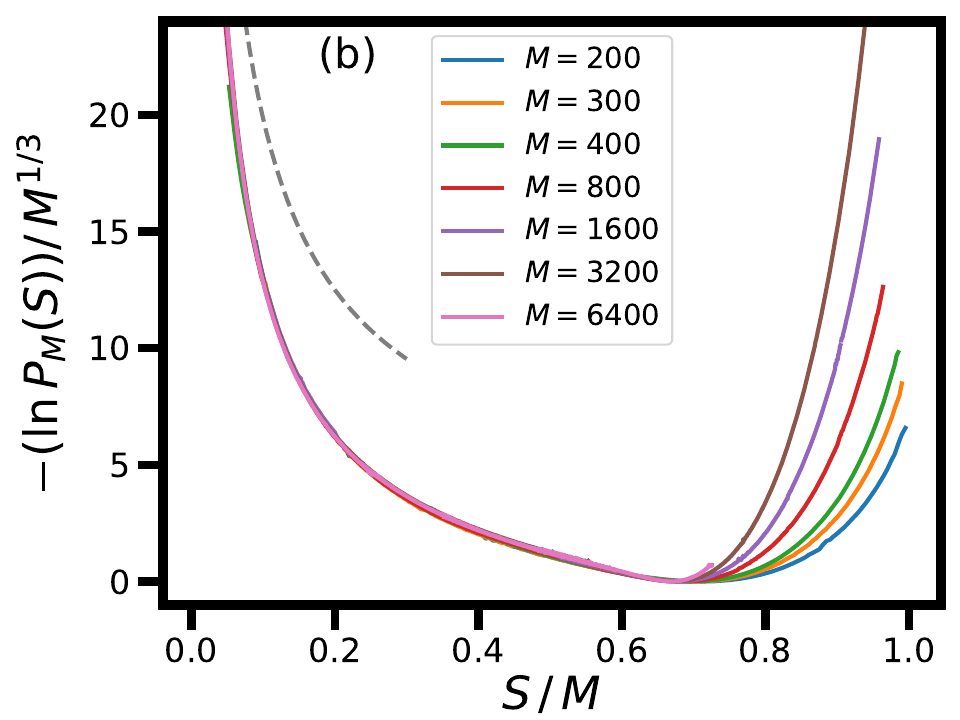}
\hspace{1mm}
\includegraphics[width=0.32\textwidth,clip=]{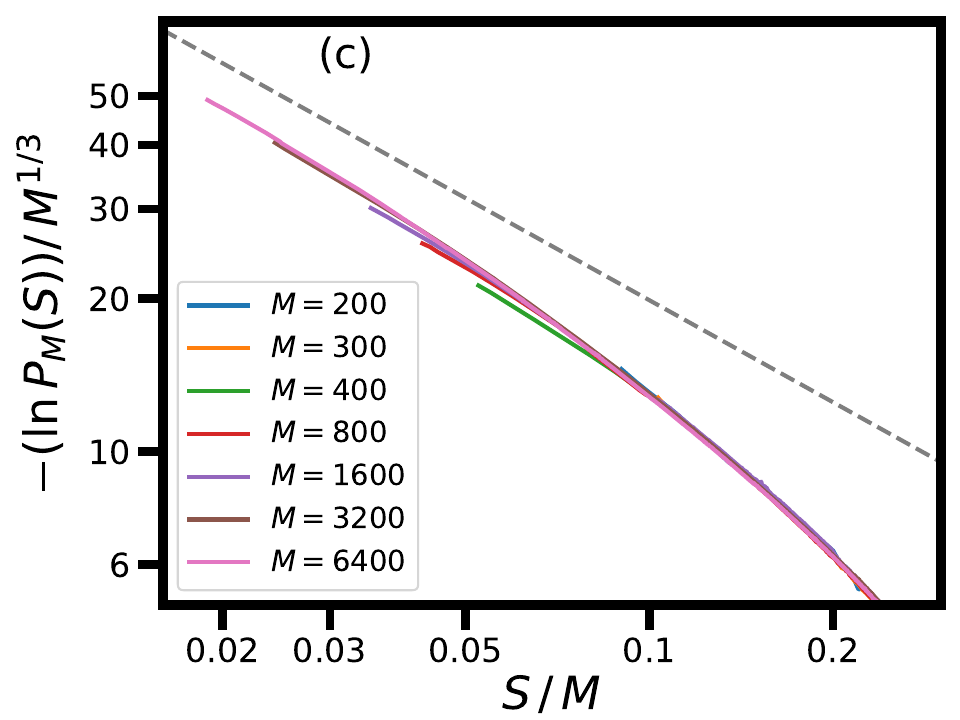}
\caption{Numerically-computed distributions $P_M(S)$ for different $M$'s, for $d=3$. The data from different $M$'s, properly rescaled, collapse onto a single curve in the right and left tails in (a) and (b) respectively. This provides numerical validation for our conjectured scaling forms, given by the LDP's \eqref{LDPRightTail} and \eqref{LDPLeftTail} with scaling functions $\Phi(a)$ and $\Psi(a)$ (respectively).
The dashed lines in (a) and (b) are the lower bound for the rate function $\Phi(a)$ given in Eq.~\eqref{PhiBounds}, and the upper bound \eqref{PsiBound} for the large deviation function $\Psi(a)$, respectively. 
The upper bound for the rate function $\Phi(a)$ given in Eq.~\eqref{PhiBounds} is far above the data that is plotted in (a); it is not visible on the scale of the figure.
In (c), the $S/M = a \ll 1$ behavior of the left tail is shown on a log-log scale. The dashed line coincides with the one plotted in (b), but at $a \ll 1$ it is expected to give the asymptotic behavior of $\Psi(a)$, see the discussion around Eq.~\eqref{PsiAsymptotic1}.
}
\label{figLDPs}
\end{figure*}

Although we are in general unable to analytically calculate the scaling functions $\Phi(a)$ and $\Psi(a)$ associated with these LDPs, we can obtain them numerically through importance sampling, see Fig.~\ref{figLDPs}. We are also able to obtain upper and lower bounds for $\Phi(a)$, see Eq.~\eqref{PhiBounds}, where the lower bound is expected to coincide with $\Phi(a)$ in the high-dimensional limit $d \gg 1$. Moreover, we obtain  an upper bound \eqref{PsiBound} for $\Psi(a)$ which we believe also yields its asymptotic behavior at $a \ll 1$, describing the far left tail of $P_M(S)$.
We also perform a (relatively straightforward) extension of our results to the case of a RW in continuous time. We find that the left tail coincides, in the leading order, with it counterpart in discrete time, while in the right tail, the scaling function for continuous time differs to that of discrete time.

\section{Large deviations of the number of distinct sites visited}
\label{sec:LargeDeviations}

\subsection{Right tail $S >\left\langle S\right\rangle $}

In the right tail, we argue that the distribution satisfies the LDP
\be
\label{LDPRightTail}
P_{M}\left(S\right)\sim e^{-M\Phi\left(S/M\right)} \, ,
\ee
in the limit $M \to \infty$ with constant $a = S/M$.
This argument is based on upper and lower bounds that we obtain for $P_{M}\left(S\right)$, both of which are exponentially small in $M$.

\smallskip

\textbf{Upper bound for $P_{M}\left(S\right)$:}
Let us denote by $N_{\text{back}}$ the number of steps that backtrack directly to the previous point, i.e.,
\be
N_{\text{back}}=\left|\left\{ t\;|\;\vect{\xi}_{t+1}=-\vect{\xi}_{t}\right\} \right| \, .
\ee
At each time step (excluding the first one), the probability of backtracking is $1/(2d)$ and the backtracking events are statistically independent of each other.
$N_{\text{back}}$ therefore follows a binomial distribution with $M-1$ trials and success probability $1/(2d)$.

An upper bound for $P_M(S)$ in the right tail comes from the observation that $N_{\text{back}}\le M-S$.
Therefore,
\be
\label{UpperBound}
\text{Prob}\left(S\ge aM\right)\le \text{Prob}\left(N_{\text{back}}\le\left(1-a\right)M\right) \, .
\ee
If $1-1/\left(2d\right)<a<1$, the latter event is exponentially unlikely in $M \gg 1$, and its probability is given by
\be
\label{UpperBound2}
\text{Prob}\left(N_{\text{back}}\le\left(1-a\right)M\right)\sim e^{-MI_{1/\left(2d\right)}\left(1-a\right)}
\ee
where
\be
I_{p}\left(y\right)=y\ln\left(\frac{y}{p}\right)+\left(1-y\right)\ln\left(\frac{1-y}{1-p}\right)
\ee
is the rate function of the binomial distribution (describing trials with success probability $p$).

\smallskip

\textbf{Lower bound for $P_{M}\left(S\right)$:}
A lower bound for $P_M(S)$ may be obtained, for any $a=S/M$, from the observation that if the final position of the RW is
\be
\vect{x}_{M}=\left(\frac{aM}{d},\dots,\frac{aM}{d}\right),
\ee
then the number of distinct points visited $S$ must be at least $aM$. It follows that
\be
\label{LowerBound}
\text{Prob}\left(S\ge aM\right)\ge \text{Prob}\left(\vect{x}_{M}=\left(\frac{aM}{d},\dots,\frac{aM}{d}\right)\right) \, .
\ee
Now, the probability on the right-hand side may be estimated by approximating  the number of steps along each of the $d$ dimensions by $M/d$ (the probability for this to occur decays as a power law in $M$ and therefore will not affect the exponential bound obtained below). Under this approximation:
(i) Each coordinate $\vect{x}_{M,i}$ of $\vect{x}_M$ is distributed according to that of a RW in $d=1$ dimension at time $M/d$, and thus
\be
\text{Prob}\left(\vect{x}_{M,i}=\frac{aM}{d}\right)\sim e^{-\frac{M}{d}I_{1/2}\left(\frac{a+1}{2}\right)} \, .
\ee
(ii) The coordinates of $\vect{x}_M$ are statistically independent of each other, and therefore,
the RHS of Eq.~\eqref{LowerBound} is  given by
\be
\label{LowerBound2}
\text{Prob}\left(\vect{x}_{f}=\left(\frac{aM}{d},\dots,\frac{aM}{d}\right)\right)\sim e^{-MI_{1/2}\left(\frac{a+1}{2}\right)} \, .
\ee

\smallskip

\begin{figure*}[ht!]
\includegraphics[width=0.48\textwidth,clip=]{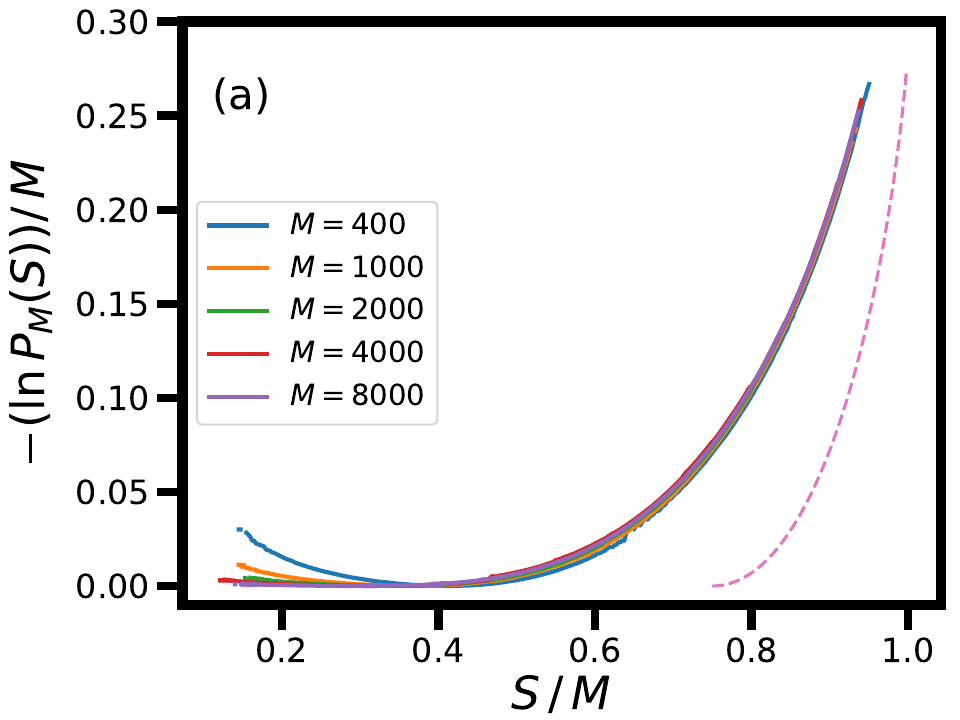}
\hspace{2mm}
\includegraphics[width=0.48\textwidth,clip=]{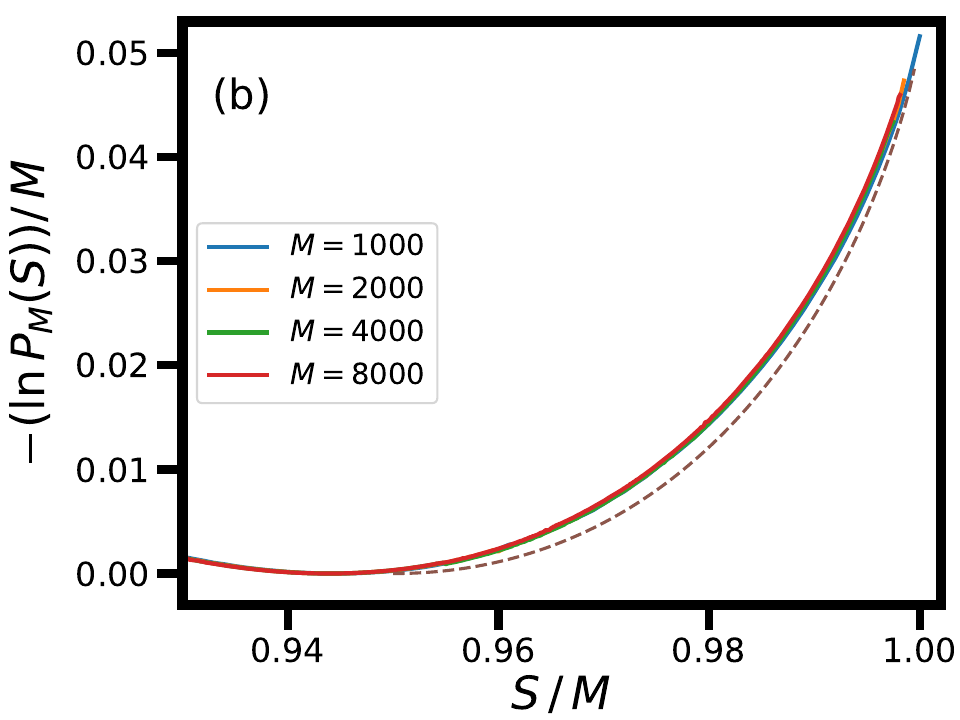}
\caption{(a) Numerical data for the right tail for $d=2$ and different values of $M$ (solid lines), together with the lower bound in Eq.~\eqref{PhiBounds} for the rate function $\Phi(a)$ (dashed line). (b) Similarly for $d=10$, only that here the dashed line plays the role not only of a lower bound, but also of the large-$d$ approximation for the rate function $\Phi(a)$, see Eq.~\eqref{IofaLarged}.
}
\label{figLDP10D}
\end{figure*}

Eqs.~\eqref{UpperBound}, \eqref{UpperBound2}, \eqref{LowerBound} and \eqref{LowerBound2} lead us to conjecture that, in the right tail, the LDP \eqref{LDPRightTail} is satisfied. Moreover, they provide bounds for the rate function
\be
\label{PhiBounds}
I_{1/\left(2d\right)}\left(1-a\right)\le\Phi\left(a\right)\le I_{1/2}\left(\frac{a+1}{2}\right) \, ,
\ee
where the lower bound for $\Phi(a)$ is only valid for $1-1/(2d) < a < 1$.
Indeed, we found that results of numerical simulations for $d=2,3,10$  appear to show excellent agreement with the LDP \eqref{LDPRightTail} over a broad range of $M$'s, see Figs.~\ref{figLDPs}(a) and \ref{figLDP10D}, and that the numerically-computed rate functions $\Phi(a)$ indeed satisfy the bounds \eqref{PhiBounds}.
Details of the numerical importance-sampling method that we used to compute $P_M(S)$ are given in Appendix \ref{app:Numerics}.

We note that the arguments given here for the right tail are valid for all dimensions $d \ge 1$. For $d>2$, we expect Eq.~\eqref{LDPRightTail} to be valid in the right tail, i.e., for all $E_d < a < 1$.
For $d \le 2$, since $\left\langle S\right\rangle $ grows slower than linearly with $M$, any value $0 < a < 1$ that is fixed in the large-$M$ limit belongs to the right tail, so we expect Eq.~\eqref{LDPRightTail} to be valid for all $0 < a < 1$.
However, for $d=2$, due to the very slow (logarithmic) convergence of $\left\langle S\right\rangle /M$ to zero as $M \to \infty$, it is in practice very difficult to test Eq.~\eqref{LDPRightTail} numerically on the full range $0<a<1$, as that would require simulations with extremely large $M$'s. We leave such a numerical test to future work.
Nevertheless, we were able to numerically verify Eq.~\eqref{LDPRightTail} for $d=2$ on the range $0.5 < a < 1$ by using moderately large $M$'s, see Fig.~\ref{figLDP10D}(a).

In the high-dimensional limit $d \gg 1$, it becomes highly unlikely for the RW to encounter any previously-visited site except by backtracking. This implies that the upper bound that we obtained for $P_M(S)$ is expected to be in fact a fairly good approximation for $P_M(S)$, since $S \simeq M-N_{\text{back}}$. This in turn implies that the rate function is expected to be approximately
\be
\label{IofaLarged}
\Phi(a) \simeq I_{1/\left(2d\right)}\left(1-a\right), \quad d\gg1.
\ee
In Fig.~\ref{figLDP10D}(b) we compare the prediction \eqref{IofaLarged} to results of numerical simulations with $d=10$ and observe good agreement.

\subsection{Left tail $S < \left\langle S\right\rangle $}

In the left tail, we obtain a simple lower bound for $P_M(S)$ for dimensions $d > 2$ (see a brief comment on the case $d=2$ at the end of this subsection). We notice that if the RW remains inside a ball with of radius $R$ (whose center can assumed, for simplicity, to be at the origin), then the number of distinct sites visited ($S$) is at most the number of lattice points $N_R$ in the ball,
$S \le N_R$.
From here it immediately follows that, for any $R$,
\be
\text{Prob}(S \le N_R) \ge P_{\text{survival}}(R;M)
\ee
where $P_{\text{survival}}(R;M)$ is the survival probability of the RW within the ball of radius $R$ (i.e., the probability that the RW stays inside the ball) up to time $M$.

Choosing $R$ such that $N_R = a M$ for some fixed value $0 < a < 1$, one finds that (at $M \gg 1$) $R \gg 1$, and therefore $N_R$ is approximately the volume of the ball,
\be
\label{NRDef}
N_R \simeq V_d R^d \, ,
\ee
where $V_d$ is the volume of the $d$-dimensional ball of unit radius,
$V_3 = 4\pi / 3$ etc.
This chosen radius $R \sim M^{1/d}$ is much smaller than the typical distance $\sim \sqrt{M}$ that the RW is expected to reach (since we assumed $d>2$). Therefore, the survival probability $P_{\text{survival}}(R;M)$ can be approximated by that of a diffusing particle, where the diffusion coefficient for the RW is $D = 1/(2d)$, inside a ball of radius $R$. The latter behaves, at long times, as
\be
\label{Psurvival}
P_{\text{survival}}(R;M)\sim e^{-c_{d}DM/R^{2}}
\ee
where $c_d$ is a $d$-dependent numerical constant, $c_3 = \pi^2$ etc, which is related to the fundamental eigenvalue of the Laplace operator inside the ball with Dirichlet boundary conditions (see e.g. \cite{TBA2016}).

Putting Eqs.~\eqref{NRDef} and \eqref{Psurvival} together, and using that in the left tail $N_R < \left\langle S\right\rangle$, one can approximate
$\text{Prob}\left(S\le N_{R}\right)\sim\text{Prob}\left(S=N_{R}\right)$,
we obtain, in the large-$M$ limit,
\be
\label{LowerBoundLeft}
P_{M}\left(S=aM\right)\ge e^{-\alpha_{d}M^{1-2/d}/a^{2/d}},\quad\alpha_{d}=\frac{c_{d}V_{d}^{2/d}}{2d} \, .
\ee

Based on this lower bound \eqref{LowerBoundLeft} for $P_M(S)$, we conjecture that the left tail of the distribution follows an LDP with anomalous scaling,
\be
\label{LDPLeftTail}
P_{M}\left(S\right)\sim e^{-M^{1-2/d}\Psi\left(S/M\right)} \, .
\ee
Moreover, it follows from \eqref{LowerBoundLeft} that the large-deviation function $\Psi$ is bounded from above by
\be
\label{PsiBound}
\Psi(a) \le \alpha_{d}/a^{2/d} \, .
\ee
In the left tail, unlike the right tail, we were not able to obtain an upper bound for $P_M(S)$.
Nevertheless, we found that Eq.~\eqref{LDPLeftTail} appears to agree with numerical simulations with finite $M$, and that the numerically-measured large-deviation function indeed satisfies the bound \eqref{PsiBound}, see Fig.~\ref{figLDPs}(b).

The argument given here is expected to be valid for $d>2$ in the left tail $0 < a < E_d$. 
This is also the domain on which $\Psi(a)$ is expected to be defined.
At $a \ll 1$, the RW must remain within the set of visited sites for a relatively long time. At $M \to \infty$ (with a fixed $a \ll 1$) this becomes increasingly unlikely, and we expect the dominant contribution to $P_M(S)$ to come from realizations in which the set of visited sites is a \emph{ball} that includes $S = aM$ points, since the probability to remain inside any other set of $S$ points is smaller than for the ball.
We therefore expect the bound \eqref{PsiBound} to give a good approximation of $\Psi(a)$ at $a \ll 1$, i.e., 
\be
\label{PsiAsymptotic1}
\Psi(a) \simeq \alpha_{d}/a^{2/d} \, , \quad a\ll1 \, .
\ee
This behavior is in good agreement with our numerical results, see Fig.~\ref{figLDPs}(c), although it was not numerically possible for us to reach values of $a \le 0.02$.

For $d=2$, if $S=aM$ then $S$ is in the \emph{right} tail (for any $0<a<1$ in the limit $M \to \infty$) so the argument given here for the scaling \eqref{LDPLeftTail} becomes meaningless. 
One may conjecture scaling forms for the left tail in $d=2$ that are different to \eqref{LDPLeftTail}, and the argument given here would then be useful for analyzing the behavior of the associated scaling functions. However, we found it difficult to test such conjectures numerically. Indeed, the logarithmic terms in Eqs.~\eqref{meanS} and \eqref{VarS} (for the case $d=2$) suggest that it would require very large $M$'s to observe convergence to any large-$M$ scaling form.

\subsection{Continuous time random walk}

Let us briefly consider how to extend our results to the case of a RW that is in continuous time (but discrete space), which hops, at a constant rate $r$, to one of its nearest neighbors. Let us choose units of time such that $r=1$.
We consider again the problem of finding the distribution $\mathcal{P}_T(S)$ of the number of distinct sites visited at time $T$.

This problem has a very simple connection to the discrete-time analogous problem. Indeed, using the law of total probability, one finds that
\be
\label{PTSexact}
\mathcal{P}_{T}\left(S\right)=\sum_{M=0}^{\infty}q_{T}\left(M\right)P_{M}\left(S\right) \, ,
\ee
where
\be
\label{Poisson}
q_{T}\left(M\right)=\frac{e^{-T}T^{M}}{M!}
\ee
is the probability that the RW performed exactly $M$ steps up to time $T$ (according to the Poisson distribution).

To calculate the large-deviation scaling form of $\mathcal{P}_{T}\left(S\right)$, we will first need to calculate that of the Poisson distribution \eqref{Poisson}. Using the Stirling formula, we obtain
\be
\ln q_{T}\left(M\right)\simeq-T+M\ln T-M\ln M+M \, ,
\ee
which, in the limit $T \to \infty$ with constant $m = M/T$, yields the LDP
\be
\label{PoissonLDP}
q_{T}\left(M\right)\sim e^{-T\alpha\left(M/T\right)},\quad\alpha\left(m\right)=1+m\ln m-m \, .
\ee
Now, plugging Eq.~\eqref{PoissonLDP} into \eqref{PTSexact}, and approximating the sum by an integral, we obtain
\be
\label{PTSintegral}
\mathcal{P}_{T}\left(S\right)\sim\int_{0}^{\infty}e^{-T\alpha\left(m\right)}P_{Tm}\left(S\right)dm \, .
\ee
Next, we use the large-deviations forms of $P_{M}\left(S\right)$ obtained above, i.e., Eq.~\eqref{LDPRightTail} and \eqref{LDPLeftTail} for the right and left tails respectively.
In the right tail, this yields a LDP
\be
\label{PTSLDPright}
\mathcal{P}_{T}\left(S\right)\sim\int_{0}^{\infty}e^{-T\left[\alpha\left(m\right)+m\Phi\left(S/\left(Tm\right)\right)\right]}dm\sim e^{-T\tilde{\Phi}\left(S/T\right)}
\ee
with a rate function given by
\be
\tilde{\Phi}\left(a\right)=\min_{0\le m<\infty}\left[\alpha\left(m\right)+m\Phi\left(a/m\right)\right] \, ,
\ee
where we used the saddle-point approximation to evaluate the integral in \eqref{PTSLDPright}.
In the left tail, however, due to the anomalous scaling in Eq.~\eqref{LDPLeftTail}, the dominant contribution to the integral \eqref{PTSintegral} comes from the vicinity of the point $m=1$, where $\alpha(m)$ vanishes. In other words, in the left tail, fluctuations (in the leading order) do not involve an atypical number of RW steps, and one simply obtains
\be
\mathcal{P}_{T}\left(S\right)\sim P_{M=T}\left(S\right)\sim e^{-T^{1-2/d}\Psi\left(S/T\right)}\,.
\ee

\section{Discussion}
\label{sec:discussion}

To summarize, we used analytic arguments to obtain the scaling behavior of large deviations of the number of distinct sites $S$ visited by a RW after $M \gg 1$ steps, in dimension $d \ge 2$. We then verified these behaviors through numerical simulations, and numerically computed the associated scaling functions. The predicted behaviors are highly asymmetric, such that large fluctuations with $S<\left\langle S\right\rangle $ are much likelier than those with $S>\left\langle S\right\rangle $.
We were able to calculate the scaling function describing the right tail of $P_M(S)$ analytically in the limit $d \gg 1$, and also to analytically obtain $P_M(S)$ in the limit $M \ll S$.
We extended our results to the case of a continuous-time RW, where we find that the behavior is very similar to the discrete-time case, but with a different rate function in the right tail.

A challenging future direction is to calculate the functions $\Phi(a)$ and $\Psi(a)$ analytically (at all $a$ and $d$), or at least to obtain asymptotic behaviors and/or bounds beyond the ones obtained here.
It would also be interesting to match the known existing results for the typical fluctuations regimes with those obtained here in the large-deviation regimes.
The case $d=2$ is particularly challenging. It would be interesting to understand the left tail behavior for this case, and to obtain more extensive numerical results, for larger $M$'s than we were able to perform here.

One could consider extensions to $N>1$ RWs \cite{LTHSW92, MT12, KMS13, MS24} or to a RW with resetting \cite{BMM22}. Additional possible extensions are to other observables that are related to $S$, such as the perimeter (in $d=2$) or surface area (in $d>2$) of the trajectory of the RW, or the number of ``islands'' of unvisited sites that it encloses \cite{WCH97}.

An alternative way to quantify the territory of a RW is through the area (in $d=2$) or (hyper-)volume (in $d>2$) of the convex hull of its trajectory \cite{Worton1995, GPH1995}. Large deviations of quantities related to the convex hull were studied numerically \cite{Claussen2015, Dewenter2016, Schawe2017, Schawe2018, Schawe2019, Schawe2020} and analytically \cite{Visotsky2021, Visotsky2023, MS2024}. Convex hulls that are much smaller than average tend to come about from realizations in which the RW stays inside a ball for a long time \cite{MS2024}, similarly to the picture presented here for the far left tail $S\ll\left\langle S\right\rangle $. On the other hand, the probability for larger-than-average convex hulls is dominated by a single most likely trajectory, and it this appears to be different to the behavior in the right tail for the problem studied in the present work.

\bigskip

\section*{ACKNOWLEDGMENTS}

NRS acknowledges support from the Israel Science Foundation (ISF) through Grant No. 2651/23, and from the Golda Meir Fellowship.

\bigskip

\appendix

\section{Details of numerical simulations}
\label{app:Numerics}
\renewcommand{\theequation}{A\arabic{equation}}
\setcounter{equation}{0}

\subsection{Basic version of the algorithm}

In order to numerically sample the full distribution $P_M(S)$ including its large-deviation tails, we employed the importance-sampling method. This algorithm has by now become a standard numerical tool for simulating rare events -- see, e.g., Ref.~\cite{Hartmann25} for a pedagogical introduction -- so we shall confine ourselves to a relatively brief description of its application to our problem.
The algorithm is based on generating a Monte-Carlo Markov chain of realizations of the RW. At each iteration, the realization may (or may not) be modified according to a Metropolis rule, as we will now explain. The goal of the algorithm is to realizations of the RW that are biased towards atypical (larger or smaller than average) values of $S$, and it employs a biasing parameter $\theta \in \mathbb{R}$.

We represent a realization by the sequence of $M$ steps taken by the RW,
$\vect{\xi}_1, \dots, \vect{\xi}_M$.
The number of distinct sites visited is a function of these steps,
$S = S(\vect{\xi}_1, \dots, \vect{\xi}_M)$.
Given a realization, the basic Metropolis step consists of choosing a random time $t\in\left\{ 1,\dots,M\right\} $, and then choosing a random step from one of the $2d$ vectors
$\tilde{\vect{\xi}}_t \in \left\{ \pm\hat{\vect{e}}_{1},\dots,\pm\hat{\vect{e}}_{d}\right\} $
(both choices are performed using a uniform distribution).
We then calculate the distinct number of steps for the modified realization,
$\tilde{S} = S(\vect{\xi}_1, \dots, \vect{\xi}_{t-1}, \tilde{\vect{\xi}}_{t}, \vect{\xi}_{t+1}, \dots, \vect{\xi}_M)$.

\begin{figure*}[ht!]
\includegraphics[width=0.48\textwidth,clip=]{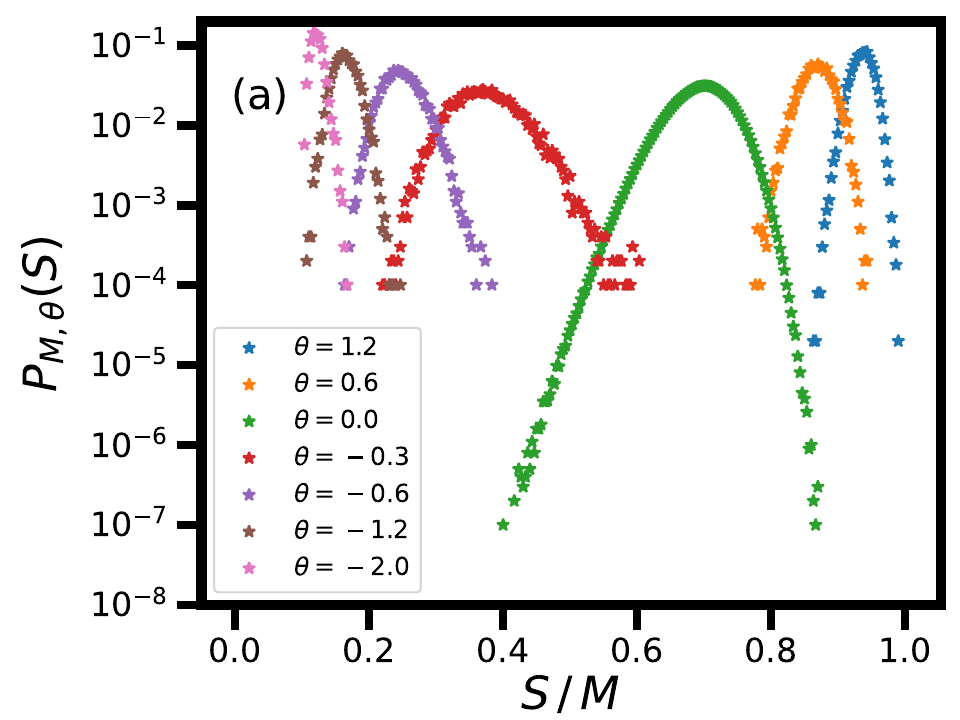}
\hspace{1mm}
\includegraphics[width=0.48\textwidth,clip=]{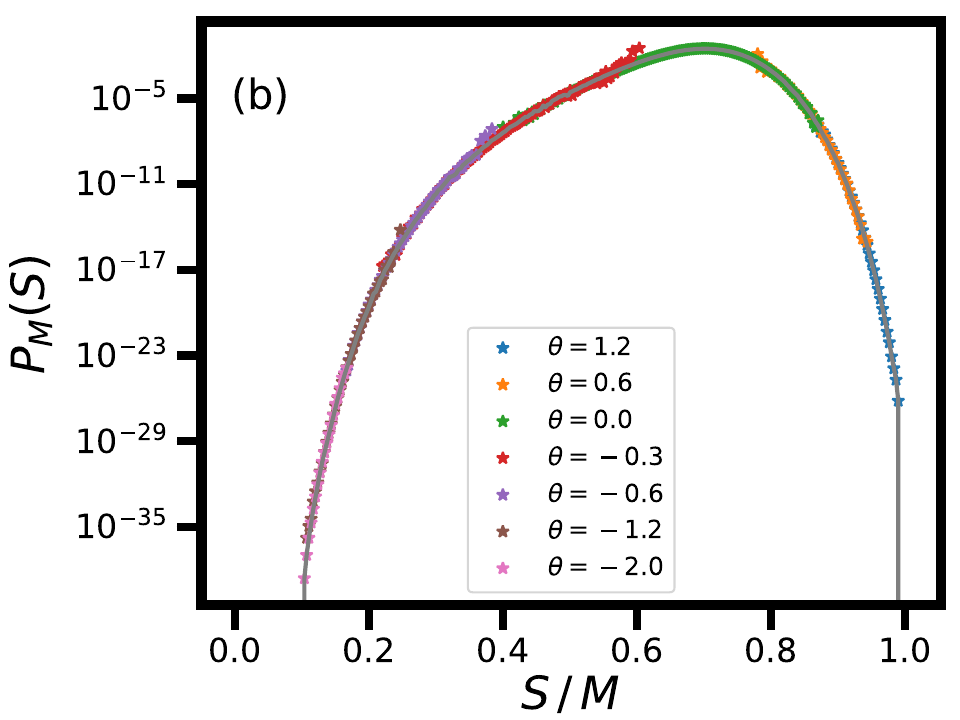}
\caption{(a) Numerically-computed biased distributions $P_{M,\theta}(S)$ for $M=300$ and $d=3$, together with the unbiased case $\theta=0$.
(b) Markers: The unbiased distribution $P_{M}\left(S\right)$ computed from the biased ones using Eq.~\eqref{PTSFromPTheta}. The different colors correspond to different values of $\theta$, which give $P_{M}\left(S\right)$ in different ranges. The normalization constants $Z(\theta)$ in \eqref{PTSFromPTheta} are estimated numerically by using the overlapping regions of different $\theta$'s, together with the known normalization $Z(\theta=0)=1$ for the unbiased case.
Solid line: the distribution $P_{M}\left(S\right)$ obtained by using all of the data from the different $\theta$'s.
}
\label{figHists300}
\end{figure*}

The Metropolis step ends by choosing whether or not to accept the change
\be
\label{MetropolisStep}
\vect{\xi}_{t} \to \tilde{\vect{\xi}}_{t},
\ee
as we shall now describe.
We calculate the quantity
$p = e^{\theta \Delta S}$,
where $\Delta S = \tilde{S} - S$ is the change in the number of distinct sites due to the proposed Metropolis step (and $\theta$ is the biasing parameter).
Now, if $p>1$, the change \eqref{MetropolisStep} is accepted. This corresponds to a change $\Delta S$ that is in the same direction as the sign of the biasing parameter.
If $0<p<1$, the change \eqref{MetropolisStep} is accepted with probability $p$.

The Metropolis dynamics described above leads to a biased, Boltzmann steady state on the space of realizations,
\be
\label{BiasedEnsemble}
P_{\theta}(\vect{\xi}_1, \dots, \vect{\xi}_M) \propto P_0(\vect{\xi}_1, \dots, \vect{\xi}_M) e^{\theta S(\vect{\xi}_1, \dots, \vect{\xi}_M)} \, ,
\ee
where $ P_0(\vect{\xi}_1, \dots, \vect{\xi}_M) = (2d)^{-M}$
is the unbiased, uniform distribution over all possible realizations.
This is easily verified by noticing that in the steady state, detailed balance is satisfied.
We now sample $S$ from realizations that are generated from the biased ensemble \eqref{BiasedEnsemble}, and this results in a biased distribution of $S$, that is given by
\be
P_{M,\theta}\left(S\right)=\frac{1}{Z\left(\theta\right)}P_{M}\left(S\right)e^{\theta S} \, ,
\ee
where $Z(\theta)$ is a normalization factor. Inverting this relation,
\be
\label{PTSFromPTheta}
P_{M}\left(S\right)=Z\left(\theta\right)P_{M,\theta}\left(S\right)e^{-\theta S} \, ,
\ee
we are then able to access different parts of the unbiased distribution $P_M(S)$ in simulations, including its tails, by choosing appropriate values of $\theta$, up to the normalization factor $Z(\theta)$. The latter is then estimated numerically by matching the results for $P_{M}\left(S\right)$ for different values of $\theta$, together with the known normalization $Z(\theta=0)=1$ for the unbiased case.

In Fig.~\ref{figHists300} we demonstrate this procedure for simulations with $d=3$ and $M=300$. The resulting $P_M(S)$ was used in Fig.~\ref{figLDPs} of the main text (and the curves plotted there for other values of $M$ were obtained similarly).
It is important to note that the algorithm described here may be used for studying the large deviations of any observable quantity, and not just the number of distinct sites visited.

\subsection{Improved version of the algorithm}

The algorithm described above performs very well to obtain the right tail of the distribution $P_M(S)$. However, in the left tail, it is not so computationally efficient: We found that, for very large $M$ (say, $M >1000$) and far into the left tail (corresponding to values of $\theta < -1.5$), it was necessary to run a huge amount of iterations until the algorithm converged.
We believe that this phenomenon occurs because, in the left tail, a single Monte-Carlo step \eqref{MetropolisStep} can change the value of $S$ by a huge amount. This in turn causes the algorithm to get jammed in configurations which give a local minimum of $S$ with respect to dynamics given by steps of the type \eqref{MetropolisStep}, since the value of $p$ is extremely small for nearly any possible step in such a configuration. Such configurations are metastable states of the (biased) Monte-Carlo dynamics, and a huge number of iterations is required to escape them.

As a remedy, we introduced another type of Monte-Carlo step, which switches between two consecutive increments of the RW, i.e.,
\be
\label{MetropolisStep2}
\vect{\xi}_{t} \leftrightarrow \vect{\xi}_{t+1}
\ee
where $t$ is chosen uniformly from $t\in\left\{ 1,\dots,M-1\right\} $.
The advantage of steps of the type \eqref{MetropolisStep2} is that they leave all of the $\vect{x}_{i}$'s unchanged, except (possibly) for $i=t$. As a result, they lead to a difference $\Delta S\in\left\{ 0,1,-1\right\} $, and this leads to values of $p$ that are not negligible.

Of course, Monte-Carlo steps of the type \eqref{MetropolisStep2} would not suffice to explore the full configuration space, since they conserve the total number of RW increments in each of the $2d$ possible directions (and in particular, they always leave $\vect{x}_M$ unchanged).
Therefore, in our simulations, at each Monte-Carlo step we propose a change of the first type \eqref{MetropolisStep} with probability $q$, and of the second type \eqref{MetropolisStep2} with probability $1-q$, where $q$ is a parameter that we usually set to the value $q=1/2$ (the change is then accepted with probability $p$ as described in the previous subsection).
We found that this version of the algorithm significantly improved our ability to sample the left tail of the distribution at large $M$'s. For instance, it enabled us to reach a minimal value of $S/M \simeq 0.025$ at $M=3200$, as opposed to $S/M \simeq 0.07$ for the basic version of the algorithm (corresponding to probabilities of order $10^{-250}$ and $10^{-120}$, respectively).

\end{document}